# Econophysics Macroeconomic Model

Victor Olkhov

TVEL, Kashirskoe sh. 49, Moscow, 115409, Russia

victor.olkhov@gmail.com

## Abstract

This paper presents macroeconomic model that is based on parallels between macroeconomic multi-agent systems and multi-particle systems. We use risk ratings of economic agents as their coordinates on economic space. Aggregates of economic or financial variables like Investment, Assets, Demand, Credits and etc. of economic agents near point $x$ define corresponding macroeconomic variables as functions of time $t$ and coordinates $x$ on economic space. Parallels between multi-agent and multi-particle systems on economic space allow describe transition from economic kinetic-like to economic hydrodynamic-like approximation and derive macroeconomic hydrodynamic-like equations on economic space. Economic or financial transactions between economic agents determine evolution of macroeconomic variables This paper describes *local macroeconomic approximation* that takes into account transactions between economic agents with coordinates near same point $x$ on economic space only and describes interaction between macroeconomic variables by linear differential operators. For simple model of interaction between macroeconomic variables as Demand on Investment and Interest Rate we derive hydrodynamic-like equations in a closed form. For perturbations of these macroeconomic variables we derive macroeconomic wave equations. Macroeconomic waves on economic space can propagate with exponential growth of amplitude and cause irregular time fluctuations of macroeconomic variables or induce economic crises.

# 1.Introduction

Econophysics during last decades made a lot for economic and financial modeling [1-8]. Financial markets and price dynamics, market trends and crises forecasting, market strategies, risk and insurance assessment use methods and models of theoretical and statistical physics. Meanwhile deep distinctions and misunderstandings between languages of economics and finance on one hand and theoretical physics on other hand [9] prevent mutual beneficial development. In this paper we try to overcome at least part of these distinctions and develop macroeconomic theory based on notions of economics and finance with help of certain parallels to kinetics and hydrodynamics.

We agree with statements [6] that direct applications of physical models and methods to economic and financial problems give no effect. Distinctions between economic and physical systems are so vital that physical methods should be completely reconstructed to be useful for economic modeling. We hope that variety of new problems that should be solved to establish economic theory in a rigorous form alike to current state of theoretical physics should be interesting to physicists. As well we do hope that our models may present new treatment of economic and financial problems and deliver benefits for econometrics, economic modeling and forecasting. Perfect treatments of existing problems in economic theories are presented in numerous works of leading economists as Morgenstern [10], Lucas [11], Lucas and Sargent [12], Sims [13], Blanchard [14], McCombie and Pike [15]. These papers should be ground for any econophysics study. We do hope that our model might respond some of them.

We develop economic theory on base of two common economic notions: economic agents and risk ratings of economic agents. Agent-based models argue decision-making of economic agents, describe trading strategies, game theories, behavioral economics, equilibrium theory [16-21]. Economic agent is a general term that describes any participant of economic or financial relations like Companies and Firms, Banks and Exchanges, Privet Investors and Households and etc. It is assumed that economic agents take rational or non rational decisions, follow personal expectations, develop and maintain its market strategy and that should explain behavior of economic agents – Producers and Consumers, Investors and Borrowers and etc.

We avoid discussions about agent's strategies and their decisions and propose regard economic agents as primary, simple units of macroeconomic *alike to* particles in kinetic models in physics. Let treat economic agents like simple economic particles that have many economic variables that describe Demand and Supply, Investment and Production Function of economic agents and so on. Such approach to agent-based models allows develop bridge between language of economic modeling and language of physics and develop parallels between description of economic multi-agent system and description of physical multi-particle system.

All macroeconomic variables are composed by corresponding variables of economic agents. Macroeconomic Demand is determined by aggregation of Demand of separate economic agents. Value Added of economic agents [22] define GDP of entire economics. Macroeconomic Consumption, Investment, Credits and Profits – all macroeconomic variables are determined by corresponding variables of separate economic agents. That is *alike to* relations between physical properties of macro system and physical properties of particles that constitute this macro system. There is no need to repeat that properties of agent-based economic system and physical multi-particle system are completely different. We outline parallels between them only.

To develop similarities between economic multi-agent system and multi-particle physical system one should define certain economic analog of space that can allow describe economic agents alike to multi-particle systems in physics. Moreover, such economic space should have origin and roots in economics and finance and adopt general economic relations and phenomena's. Usage of space in economics usually refers to spatial economics [23, 24], but that approach is helpless. To develop parallels between multi-agent and multi-particle systems we introduce economic space notion [25, 26] that reflects internal economic properties and allows develop general frame to macroeconomic and macro finance modeling. Economic space approach gives new look on option pricing theory and we derived generalization of Black-Scholes-Merton equations on $n$-dimensional economic space [25, 26, 27]. In this paper we describe macroeconomic multi-agent systems on $n$-dimensional economic space. We develop parallels to kinetics and hydrodynamics nevertheless phenomena's of economic agent systems have nothing common with nature of kinetics and hydrodynamics of multi-particle physical systems.

We propose use risk ratings of economic agents as their coordinates on economic space. Risk ratings of economic agents are provided for decades by international rating agencies as Fitch's [28], S&P [29], Moody's [30]. Risk agencies define risk ratings of huge corporations and banks. These ratings are used as measure of assets security, sustainability and helps take investment decisions, estimate market prices of assets and etc. Risk ratings take finite number of values or risk grades. Let propose that risk ratings can be measured for all economic agents. Let assume that risk grades can be discreet as it is now or can be continuous $R$. Let call discreet or continuous space of risk grades as economic space. Let propose that simultaneous assessment of $n$ different risks determine coordinates of economic agents on $n$-dimensional economic space. These assumptions allow describe macroeconomics *alike to* description of multi-particle systems and derive hydrodynamic-like equations on economic space.

Our macroeconomic model is pure theoretical as no econometric data required to verify predictions of our theory exist. Current risk ratings data are not sufficient develop macroeconomic models on economic space. We do hope that required enhancement of risk assessments, econometric observations and data performance can improve economic modeling, forecasting and management.

The rest of the paper is organized as follows: in Section 2 we introduce economic space and discus related economic and physical problems. In Section 3 and 4 we present kinetic-like and hydrodynamic-like economic models. In Section 5 we study simple interactions between two macroeconomic variables that allow derive hydrodynamic-like equations in a closed form and derive wave equations on macroeconomic perturbations alike to acoustic equations in fluids. Simple example in Section 6 demonstrates how waves can induce time fluctuations of macroeconomic variables. Derivation of acoustic-like wave equations is not too interesting for theoretical physics papers but as we know it is the first evidence and description of wave processes in macroeconomic and financial models. In Section 7 we argue diversity of macroeconomic models and some open problems that can be interesting for physicists. Conclusions are in Section 8.

## 2. Definition of Economic space

Description of macroeconomic multi-agent system alike to multi-particle system requires introduction of economic analogy of space that allows define coordinates of economic agents *alike to* coordinates of physical particles. As analogy coordinates we suggest risk ratings of economic agents [25, 26]. Here we present brief reasons for economic space definition.

International rating agencies [28-30] estimate risk ratings of economic agents as Banks and Corporations, Firms and Enterprises. Risk ratings take values of risk grades and noted as *AAA, BB, C* and so on. Let treat risk grades like *AAA, BB, C* as points $x_1, x_2,... x_m$ of discreet space. Let propose, that risk assessments methodologies can estimate risk ratings for all agents of entire economics: for huge Banks and for small households. If so, risk ratings distribute all economic agents of the entire economics over points of finite discreet space determined by set of risk grades. There are a lot of different risks those impact economic processes. Let regard grades of single risk as points of one-dimensional space and simultaneous assessments of *n* different risks as coordinates of economic agent on *n*-dimensional space. Let propose, that risk assessments methodologies can be generalized in such a way that risk grades can fill continuous space *R*. Thus risk grades of *n* different risks establish $R^n$.

Let define economic space as any mathematical space that map economic agents by their risk ratings as space coordinates. Number *n* of risks ratings measured simultaneously determines dimension *n* of economic space. Let put positive direction along each risk axis as risk growth direction. Let assume that all economic agents of entire economics are "independent" and sum of extensive (additive) economic variables of any subset of agents equals economic variable of the entire subset. For example, sum of Assets of any two economic agents equal their collective Assets. Let assume that econometric data contains data about risk ratings and all economic variables of each economic agent. These assumptions require significant development of current econometrics and statistics. Quality and granularity of current U.S. National Income and Product Accounts system [22] gives hope that all these problems can be solved.

Definition of economic space as grades of all economic and financial risks arises additional tough problem. There are a plentiful number of different economic and financial risks but their influence on macroeconomic evolution is very different. There are many risks that induce small influence and their action on entire macroeconomics might be neglected. We assume that current macroeconomic dynamics should be determined by action of one-two-three major risks and assessment of their ratings should define economic space. Thus definition of economic space $R^n$ requires selection of *n* risks with major impact on economic agents and macroeconomic processes. These *n* risks define initial state of economic space $R^n$. Selection of most valuable risks requires procedures that allow measure and compare influence of different risks on entire economics and economic agents. Assessment and comparison of different risks and their influence on economic agents establish tough problems and such models should be developed. Risk assessments methodologies and procedures, comparison of risk influence on performance of economic agents and on macroeconomic dynamics can establish procedures *alike to* physical measurement theory and measurement procedures. It allows develop relations between economic theory and econometric statistics *alike to* interdependence between physical theory and experimental measurements. Solution of this hard problem requires close collaboration between physicists and economists.

Economic and financial risks have random nature and can unexpectedly arise and then vanish. Thus some current major risks that define initial representation of economic space $R^n$ can accidentally disappear and other major risks may come to play. Thus economic space representation can be changed randomly. Description of economic dynamics and forecasting for time term *T* requires prediction of *m* main risks that can play major role in a particular time term and can define economic space $R^m$. Such set of *m* risks determine target state of economic space $R^m$. Transition of economic modeling on initial economic space $R^n$ to target economic space $R^m$ requires description of decline of action of initial set of *n* risks on entire economics and description of growth of influence of new *m* risks. Such stochastic scenarios are completely different from physical models that study complex dynamics of random fields and particles determined on constant physical space.

Current macroeconomics describes relations between macroeconomic variables as Demand and Supply, Production Function and Investment, Economic Growth and Consumption each treated as function of time. Introduction of economic space gives ground for definition of macroeconomic variables as functions of time and *coordinates*. This small step opens doors for wide application of mathematical physics methods and models that should be transformed to adopt economic and financial phenomena's.

Below we present economic model on economic space $R^n$ in the assumption that economic agents are under action of constant set of *n* major risks. We describe macroeconomics *alike to* kinetics and hydrodynamics and derive wave-like equations for macroeconomic variables. Up now notions of waves in economics and finance are used to describe Kondratieff waves [31], inflation waves, crisis waves, etc. All these issues don't describe any waves but time oscillations of economic variables only. Description of waves requires space and introduction of economic space notion gives ground for development of economic wave theory.

## 3. Macroeconomic kinetics

Let treat macroeconomics as set of economic agents on economic space. Economic agents can move on economic space *alike to* particles. For convenience let call economic agents as economic particles or e-particles and economic space as e-space. Let assume that each e-particle is described by $l$ extensive (additive) economic variables $(u_1,...u_l)$ as Supply and Demand, Production Function and Capital, Consumption and Value and etc. Let study macroeconomics on e-space $R^n$ that reflects action of constant set of $n$ risks. Risk ratings of e-particles play role of their coordinates on e-space $R^n$.

Each e-particle on e-space $R^n$ at moment $t$ is described by coordinates $x=(x_1,...x_n)$, velocity $v=(v_1,...v_n)$, and extensive economic variables $(u_1,...u_l)$. Extensive economic variables of economic agents are additive and admit averaging by probability distributions. Intensive economic variables, like Prices or Interest Rates, cannot be averaged directly. Enormous number of extensive variables like Value and Capital, Demand and Supply, Profits and Savings, Consumption and Investment etc., describe economic and financial performance of each e-particle and are origin of extreme complexity of economic systems. As usual, macroeconomic variables are defined as aggregates of corresponding values of all economic agents of entire macroeconomics. For example, Demand of entire macroeconomics equal aggregate Demand of all economic agents and GPD can be calculated as aggregate Value Added of all economic agents [22]. Let introduce macroeconomic variables as aggregates of corresponding values of economic agents with coordinates $x$ on e-space.

Let assume that there are $N(x)$ e-particles at point $x$. Let state that velocities of e-particles at point $x$ equal $v=(v_1,...v_{N(x)})$. Let describe economics that has $l$ macroeconomic variables and hence each e-particle has $l$ economic variables $(u_1,...u_l)$. Let assume that values of economic variables equal $u=(u_{1i},...u_{li})$, $i=1,..N(x)$. Each extensive economic variable $u_j$ at point $x$ defines macroeconomic variable $U_j$ as sum of economic variables $u_{ji}$ of $N(x)$ e-particles at point $x$

$$U_j = \sum_i u_{ji} \; ; \quad j = 1,..l; \quad i = 1,...N(x)$$

For each macroeconomic variable $U_j$ let define analogy of impulses $P_j$ as

$$P_j = \sum_i u_{ji} v_i; \quad j = 1,..l; \quad i = 1,...N(x)$$

Let follow Landau and Lifshitz [33] and introduce economic distribution function $f=f(t,x;U_1,..U_l, P_1,..P_l)$ on *n*-dimensional e-space that determine probability to observe macroeconomic variables $U_j$ and impulses $P_j$ at point $x$ at time $t$. $U_j$ and $P_j$ are determined by corresponding values of e-particles that have coordinates $x$ at time $t$. They take random values at point $x$ due to random behavior of e-particles on e-space. Averaging of $U_j$ and $P_j$ within distribution function $f$ allows establish transition from approximation that takes into account economic variables of separate e-particles to hydrodynamic-like approximation of macroeconomics that neglect e-particles granularity. Let define macroeconomic density function $U_j(t,x)$

$$U_j(t, x) = \int U_j\, f(t, x, U_1, ... U_l, P_1, .. P_l)\, dU_1.. dU_l dP_1.. dP_l \qquad (1)$$

and impulse density $P_j(t,x)$ as

$$P_j(t, x) = \int P_j\, f(t, x, U_1, ... U_l, P_1, .. P_l)\, dU_1.. dU_l dP_1.. dP_l \qquad (2)$$

That allows define e-space velocity $v_j(t,x)$ of density $U_j(t,x)$ as

$$U_j(t, x) v_j(t, x) = P_j(t, x) \qquad (3)$$

Densities $U_j(t,x)$ and impulses $P_j(t,x)$ are determined as mean values of aggregates of corresponding economic variables of separate e-particles with coordinates $x$. Functions $U_j(t,x)$ can describe macroeconomic e-space density of Demand and Supply, Assets and Debts, Production Function and Value Added and so on. Usage of distribution function $f=f(t,x;U_1,..U_l, P_1,..P_l)$ allows describe any statistical moments of macroeconomic variables like $<U_j^m>$, correlations between economic variables $<U_jU_i>$ and so on. Operators $<..>$ define averaging by distribution function $f$. We use (1-3) as tool to establish description of densities $U=(U_1,...U_l)$, $U_j=U_j(t,x)$ as functions on e-space $R^n$ and develop macroeconomic model *alike to* hydrodynamics.

## 4. Macroeconomic hydrodynamics

Each macroeconomic density $U_1(t,x),... U_l(t,x)$ like Assets and Investment, Demand and Supply play role *similar to* fluid density $\rho(t,x)$ in physical hydrodynamics. Such analogy allows call $U_j(t,x)$ as densities of economic fluids or e-fluids. Extensive economic variables $U_j$ of e-particles define corresponding amount of e-fluids $U_j(t,x)$. Such hydrodynamic-like approximation of macroeconomics describes interactions between e-fluids $U_1(t,x),...U_l(t,x)$ and outlines parallels to multi-fluids hydrodynamics. Parallels between physical and macroeconomic densities permit obtain e-fluid equations similar to Continuity Equation and Equation of Motion for hydrodynamics [34].

Continuity Equation on macroeconomic density $U_i(t,x)$, $i=1,..l$ takes form

$$\frac{\partial U_i}{\partial t} + div(\boldsymbol{v}_i U_i) = Q_1 \qquad (4)$$

$\boldsymbol{v}_i(t,\boldsymbol{x})$ - is velocity of e-fluid $U_i$ on e-space. Left side describes the flux of density $U_i(t,\boldsymbol{x})$ through the unit volume surface on e-space $R^n$ and right hand side $Q_1$ describes factors that change density $U_i(t,\boldsymbol{x})$. Macroeconomic density $U_i(t,\boldsymbol{x})$ can change in time and during motion of the selected volume on e-space due to economic reasons. For example macroeconomic Demand in unit e-space volume can increase in time due to economic growth and fall down due to economic crisis. As well, Demand density can decrease if unit volume moves in the direction of risk growth. Integral of Demand density over e-space determines Demand of entire macroeconomics and it changes its value in time due to phases of economic cycles. Equation of Motion for macroeconomic density $U_i(t,\boldsymbol{x})$ takes form

$$U_i \left[\frac{\partial \boldsymbol{v}_i}{\partial t} + (\boldsymbol{v}_i \cdot \nabla)\boldsymbol{v}_i\right] = \boldsymbol{Q}_2 \qquad (5)$$

Left side describes the flux of $\boldsymbol{P}_i(t,\boldsymbol{x}) = U_i(t,\boldsymbol{x})\boldsymbol{v}_i(t,\boldsymbol{x})$ through the surface of unit volume on e-space, taking into account Continuity Equation. The right hand side $\boldsymbol{Q}_2$ describes factors that induce changes of macroeconomic density and velocity.

Economic and financial transactions between e-particles define evolution of densities of macroeconomic variables. This paper presents *local model* of economic and financial transactions between e-particles on e-space that takes into account transactions between e-particles with nearly same coordinates only. *Local model* of economic and financial transactions between e-particles simplifies description of macroeconomics. That assumption allows describe dynamics of macroeconomic densities *alike to* modeling collisions between e-particles and describe factors $Q_1$ and $\boldsymbol{Q}_2$ by linear differential operators on *conjugate* macroeconomic densities and their velocities. We define *conjugate* densities below and use this assumption in the next Section.

To determine right hand factors $Q_1$ and $\boldsymbol{Q}_2$ let outline that same economic variables of different e-particles do not interact with each other. For example, Supply of e-particle 1 does not depend on the Supply of e-particle 2, but depends on other economic variables like Demand, Investment and so on. As well Consumption of e-particle 1 does not depend on Consumption of other e-particle, but is determined by Income, Savings, Inflation and etc. Let state that economic variables of different e-particles do not interact and do not depend on same variables. Let neglect any interaction between same economic variables of different e-particles. That causes lack of any self-interaction of macroeconomic densities $U_i(t,x)$ and we state no economic parallels to such physical factors as pressure or viscosity. Let state that right hand side factors $Q_1$ and $Q_2$ in Continuity Equation and Equation of Motion for particular macroeconomic density $U_i(t,x)$ do not depend on any factors determined by same variables $U_i(t,x)$ but depend on economic densities $U_j(t,x)$, $v_j(t,x)$ different from $U_i(t,x)$, $v_i(t,x)$.

Let call variables $U_j(t,x)$, $v_j(t,x)$ that determine $Q_1$ and $Q_2$ factors in right hand side of hydrodynamic-like equations on variables $U_i(t,x)$ and $v_j(t,x)$ as variables *conjugate* to $U_i(t,x)$ or *conjugate* e-fluids. For example Supply may has *conjugate* variables like Demand, Investment, Credits and their velocities. Demand may *conjugate* to Supply and vice versa. Let state, that *conjugate* variables or *conjugate* e-fluids define right hand side of Continuity Equation and Motion Equations (4,5). Factors $Q_1$ and $Q_2$ in equations (4,5) describe action of *conjugate* e-fluids. Two *conjugate* e-fluids model is a simplest case that allows derive equations (4,5) in a closed form. Let study this model and possible $Q_1$ and $Q_2$ factors to obtain equations on two *conjugate* e-fluids in a closed form. As we show below, equations on two *conjugate* e-fluids allow derive wave equations on e-fluid densities perturbations. Existence of wave propagation of macroeconomic variable perturbations on e-space gives new look on macroeconomic modeling and description of economic shocks and their consequences.

## 5. Two conjugate e-fluids model

Hydrodynamic-like Eq. (4,5) describe dynamics of e-fluids on e-space for given factors $Q_1$ and $Q_2$. Let study example of two *conjugate* e-fluids model and show possible advantages of economic hydrodynamic-like approximation. Let study relations between Investment and Interest Rates. Above we call macroeconomic density $U_2(t,x)$ or e-fluid $U_2(t,x)$ as *conjugate* to e-fluid $U_1(t,x)$ if e-fluid $U_2(t,x)$ or it's velocity $v_2(t,x)$ determine factors $Q_1$ and $Q_2$ in the right hand side of Eq. (4,5) on e-fluid $U_1(t,x)$ and it's velocity $v_1(t,x)$. Factors $Q_1$ and $Q_2$ can be determined by one, two or many different e-fluids $U_j(t,x)$ and that makes macroeconomic modeling on e-space a very complex problem even in *local approximation*. There are two ways to use equations (4.5):

1. Study evolution of selected macroeconomic density $U(t,x)$ for factors $Q_1$ and $\mathbf{Q}_2$ determined by given functions of *conjugate* e-fluids. That allows describe dynamics of macroeconomic e-fluid $U(t,x)$ and it's velocity $v(t,x)$ in the given macroeconomic environment. All *conjugate* e-fluids are exogenous variables for e-fluid $U(t,x)$ and one solves equations (4,5) that describe behavior of endogenous variable $U(t,x)$ and $v(t,x)$ for given right side factors $Q_1$ and $\mathbf{Q}_2$.

2. Study equations (4,5) in the assumption that e-fluids are *self-conjugate*. For example, e-fluid $U_2(t,x)$ is *conjugate* to e-fluid $U_1(t,x)$ and vice versa. So, factors $Q_1$ and $\mathbf{Q}_2$ for equations (4,5) on e-fluid $U_1(t,x)$ are determined by e-fluid $U_2(t,x)$ and factors $Q_1$ and $\mathbf{Q}_2$ for equations on e-fluid $U_2(t,x)$ are determined by e-fluid $U_1(t,x)$. Such model allows obtain hydrodynamic-like equations on e-fluids $U_1(t,x)$ and $U_2(t,x)$ in a closed form.

Both approaches to Eq.(4,5) allow study macroeconomic models on e-space. Let study second case and derive self-consistent equations for simplest model of two *self-conjugate* e-fluids interactions. Such assumption simplifies the problem and allows study mutual relations between macroeconomic variables $U_1(t,x)$ and $U_2(t,x)$.

*5.1. Model: Demand on Investment - Interest Rate*

Let study simple model that describe well-known relations between Demand on Investment and Interest Rate. Rise in Demand on Investment lead to Interest Rate growth. Interest Rate growth induce decline of Demand on Investment. Let neglect all other factors that have influence on Investment Demand and Interest Rate and simplify relations between core macro financial variables to obtain equations (4,5) in a closed form. Demand on Investment $U_I(t,x)$ is extensive variable and Interest Rate $ir(t,x)$ is intensive economic variable. As we mentioned above, one can apply averaging procedure (1-3) to extensive (additive) variables of e-particles only. Intensive macroeconomic variables are determined as proportionality factor for relations between two extensive macroeconomic variables. Thus macroeconomic Interest Rate $ir(t,x)$ determine proportionality factor between Cost of Investment $U_C(t,x)$ and Funds $U_F(t,x)$ available for Investment. For fixed value of $U_F(t,x)$, Cost of Investment $U_C(t,x)$ for fixed time term equals:

$$U_C(t,x) = ir(t,x)\, U_F(t,x)$$

Thus for constant Funds $U_F(t,x)$ available for Investment, Cost of Investment $U_C(t,x)$ is proportional to Interest Rate $ir(t,x)$ only. Rise in Investment Demand $U_I(t,x)$ lead to growth of Interest Rate $ir(t,x)$ and that induce Cost of Investment $U_C(t,x)$ growth. Growth Cost of Investment $U_C(t,x)$ induced by rise of Interest Rate $ir(t,x)$ imply decline of Demand for Investment $U_I(t,x)$. Let replace Interest Rate $ir(t,x)$ as intensive variable by Cost of Investment $U_C(t,x)$ as extensive variable taking into account that $U_C(t,x)$ depends on Interest Rate $ir(t,x)$ only with Funds $U_F(t,x)$ being constant. That establish a model with two interacting *conjugate* e-fluids Demand for Investment $U_I(t,x)$ - $U_C(t,x)$ Cost of Investment.

Due to above assumption that economic and financial transactions are *local*, density $U_I(t,x)$ at point $x$ is determined by *conjugate* variables $U_C(t,x)$ and can be described by differential operators. Let study simples case that describes interaction of *conjugate* variables by operators *div* and *grad*. Let assume that $Q_1$ factor in the right hand side of Continuity Equation (4) on Demand for Investment density function $U_I$ describe *local* action on Cost of Investment and is proportional to divergence of velocity $v_C$ of Cost of Investment $Q_1 \sim \alpha_C \nabla \cdot v_C$. Positive divergence of Cost of Investment velocity $\nabla \cdot v_C > 0$ describes growth of supply flux with the same Cost of Investment density function $U_C$ and that increase Demand for Investment $U_I$, thus $\alpha_C > 0$. Let assume that $Q_1$ factor for Continuity Equation (4) on Cost of Investment density function $U_C$ is proportional to divergence of velocity $v_I$ of Investment Demand: $Q_1 \sim \alpha_I \nabla \cdot v_I$. Positive divergence of Demand for Investment velocity $\nabla \cdot v_I > 0$ describe source of demand flux and that increase Cost of Investment and $\alpha_I > 0$. Let state that $Q_2$ factor for Equation of Motion (5) for Demand on Investment velocity $v_I$ is proportional to gradient of Cost of Investment density $\nabla U_C$:

$$Q_2 \sim \beta_C \nabla U_C$$

Demand for Investment velocity $v_I$ can decrease in the direction of positive gradient of Cost of Investment density $U_C$. Flux of expensive Cost of Investment proposals will decline velocity of Demand for Investment flow and $\beta_C < 0$. Let state that $Q_2$ factor for Equation of Motion (5) for Cost of Investment velocity $v_C$ is proportional to gradient of Investment density $\nabla U_I$:

$$Q_2 \sim \beta_I \nabla U_I$$

Our assumptions means that Cost of Investment velocity $v_C$ increase in the direction with positive gradient of Investment density $U_I$. Indeed, Cost of Investment flow is directed in the domain with higher Demand for Investment and $\beta_I > 0$. Thus our assumptions give simple models of mutual dependence between Demand for Investment $U_I$ and Cost of Investment $U_C$ on e-space. We remind that in our model Cost of Investment depends on Interest Rate *ir(t,x)* only. Thus equations (4,5) for two *self-conjugate* e-fluids Demand for Investment $U_I(t,x)$ and Cost of Investment $U_C(t,x)$ that depends on Interest Rate *ir(t,x)* only take form:

$$\frac{\partial U_I}{\partial t} + \nabla \cdot (\boldsymbol{v}_I U_I) = \alpha_C \nabla \cdot \boldsymbol{v}_C \; ; \quad \frac{\partial U_C}{\partial t} + \nabla \cdot (\boldsymbol{v}_C U_C) = \alpha_I \nabla \cdot \boldsymbol{v}_I \tag{6.1}$$

$$U_I \left[\frac{\partial \boldsymbol{v}_I}{\partial t} + (\boldsymbol{v}_I \cdot \nabla)\boldsymbol{v}_I\right] = \beta_C \nabla U_C \; ; \quad U_C \left[\frac{\partial \boldsymbol{v}_C}{\partial t} + (\boldsymbol{v}_C \cdot \nabla)\boldsymbol{v}_C\right] = \beta_I \nabla U_I \tag{6.2}$$

$$\alpha_I > 0 \; ; \; \alpha_C > 0 \; ; \; \beta_I > 0 \; ; \; \beta_C < 0 \; ;$$

*5.2. Wave equations on e-fluid densities perturbations*

To derive macroeconomic wave equations on base of (6.1-6.2) let take small perturbations $q_I$ of Demand on Investment $U_I$ and small perturbations $q_C$ of constant Cost of Investment densities $U_C$ and assume that velocities $\boldsymbol{v}_I$ and $\boldsymbol{v}_C$ are small. Let put:

$$U_I = U_{I0} + q_I \; ; \; U_C = U_{C0} + q_C \tag{7.1}$$

and assume that derivations of $U_{I0}$ and $U_{C0}$ by time and coordinates in Eq.(6.1-6.2) are small to compare with similar derivations of $q_I$, $q_C$, $\boldsymbol{v}_I$ and $\boldsymbol{v}_C$ so we can neglect derivations by $U_{I0}$ and $U_{C0}$. In hydrodynamics similar approximations are used to derive acoustic wave equations [34]. Continuity Eq.(6.1) on small perturbations $q_{I,C}$ in linear approximation:

$$\frac{\partial q_I}{\partial t} + U_{I0} \nabla \cdot \boldsymbol{v}_I = \alpha_C \nabla \cdot \boldsymbol{v}_C \; ; \quad \frac{\partial q_C}{\partial t} + U_{C0} \nabla \cdot \boldsymbol{v}_C = \alpha_I \nabla \cdot \boldsymbol{v}_I \tag{7.2}$$

Equations of Motion in linear approximation:

$$U_{I0} \frac{\partial \boldsymbol{v}_I}{\partial t} = \beta_C \nabla q_C \; ; \quad U_{C0} \frac{\partial \boldsymbol{v}_C}{\partial t} = \beta_I \nabla q_I \tag{7.3}$$

Derivation of equations on $q_I, q_C$ from (7.1-7.3) is very simple and we omit it here:

$$\left[\frac{\partial^4}{\partial t^4} - a\Delta \frac{\partial^2}{\partial t^2} + b\Delta^2\right] q_{I,C} = 0 \tag{8.1}$$

$$a = (\alpha_I \beta_C + \alpha_C \beta_I) \; ; \; b = \beta_C \beta_I (\alpha_I \alpha_C - 1)$$

It is easy to show that for $a^2 > 4b$ there exist two positive $c^2_{1,2} > 0$ and Eq.(8.1) take form of bi-wave equations

$$\left(\frac{\partial^2}{\partial t^2} - c_1^2 \Delta\right)\left(\frac{\partial^2}{\partial t^2} - c_2^2 \Delta\right) q_{I,C} = 0 \tag{8.2}$$

Bi-wave equations (8.2) describe propagation of waves *q=q(x-ct)* with speed *c* equals $c_1$ or $c_2$ as in the direction of risks growth as in the direction of small risks. If coefficients $a^2 < 4b$ then equation (8.1) admits wave solutions with amplitudes amplification in time as exponent. So, small perturbations of Cost of Investment may induce waves that propagate on e-space with exponential growth of amplitudes:

$$q_C = \cos(\boldsymbol{k} \cdot \boldsymbol{x} - \omega t) \exp(\gamma t)$$

For $\gamma>0$ the solution will grow up and for $\gamma<0$ will dissipate.

$$\omega^2 = k^2 \frac{\sqrt{4b + 3a^2} + 2a}{8} > 0 \; ; \; \gamma^2 = k^2 \frac{\sqrt{4b + 3a^2} - 2a}{8} > 0$$

These examples demonstrate possible exponential amplification or dissipation of wave amplitudes of small macroeconomic perturbations in model of two interacting *conjugate* e-fluids. Derivation of above results is simple and we omit it here. Nevertheless even for simple model equations on disturbances of economic densities take form of bi-wave Eq.(8.2) and Green function for such bi-wave equations equals convolution of Green functions of two wave equations. Thus even simplest response on δ-function shock in economics is more complex then in physics. Existence of wave processes on e-space allows describe macroeconomic wave generation, propagation and interaction as possible wave response on macroeconomic shocks. Amplitudes amplifications of macroeconomic perturbations waves may model macroeconomic and financial crises evolution.

## 6. Time fluctuations of macroeconomic variables

All economic variables follow time fluctuations and often these fluctuations are called waves [31, 35, 36]. Meanwhile all these "waves" are only fluctuations of economic variables in time. Nature of waves requires space where these waves can propagate. Macroeconomic models on e-space uncover existence of wave equations for macroeconomic variables and start studies of economic wave generation, propagation and interaction on e-space. Above we derive simple Demand on Investment - Interest Rate interaction model that admit wave equations and present simple wave solutions. Let show that even simple waves can cause irregular time fluctuations macroeconomic variables.

Due to definition of e-space in Section 2 coordinates of e-particles define their risk ratings. Let consider simplest e-space $R$. Let assume that risk ratings of e-particles are reduced by minimum $X_{min}$ and maximum $X_{max}$ risk grades. For simplicity let take $X_{min}=0$ and $X_{max}= X$ and coordinates $x$ of e-particles on e-space $R$ follow

$$0 \leq x \leq X \tag{9.1}$$

Relation (9.1) defines simplest macroeconomic domain on e-space. Due to (7.1) macroeconomic density $U_I$ is presented as

$$U_I(t,x) = U_{I0} + q_I \quad (9.2)$$

$U_{I0}$ is constant or its derivatives are small to compare with derivatives of perturbations $q_I$. Let take simplest wave solution $q_I$ of equation (8.1; 8.2) on e-space $R$ as

$$q_I(t,x) = \cos(k \cdot x - \omega t) \quad ; \quad \omega_{1,2}^2 = k^2 c_{1,2}^2 \quad (9.3)$$

As we mentioned above, integral of macroeconomic density over e-space gives corresponding macroeconomic variable of entire economics as function of time. So, integral of Demand on Investment density $U_I(t,x)$ over e-space gives macroeconomic Demand on Investment $U_I(t)$. For assumption (9.1) integral $q_I(t,x)$ of (9.2, 9.3) by coordinate $x$ on e-space $R^2$ gives

$$U_I(t) = U_0 + q_I(t) \quad ; \quad U_0 \sim U_{I0} X \quad (9.4)$$

$$q(t) = \frac{2}{k}\sin\left(\frac{k}{2}X\right)\cos(\frac{k}{2}X - \omega_{1,2}t) \quad (9.5)$$

Due to (9.5) macroeconomic Investment $U_I(t)$ follows time oscillations with frequency $\omega$. For fixed wave speed $c^2_{1,2}$ linear equations (8.1, 8.2) may have wave solutions with random $k$ and random frequency $\omega$ that satisfy (9.3). Hence random wave vectors $k$ may induce random time oscillations of macroeconomic variables. Relations between variables like GDP, Investment, Supply and Demand etc., of entire economics treated as functions of time can be determined by complex interaction of *conjugate* macroeconomic variables. Macroeconomic density perturbations waves on e-space may be origin of time oscillations of macroeconomic variables, origin of business cycles etc.

## 7. Diversity of macroeconomic models and open problems

Macroeconomics is a very complex system and enormous number of economic variables and properties describe it state and evolution. That causes diversity and complexity of mutual dependence between macroeconomic variables. This paper presents simple model relations between macroeconomic Demand for Investment and Interest Rate. It is obvious that different macroeconomic densities can depend on *conjugate* variables in a different from. For example, dependence of Demand for Investment on Interest Rate can be different from dependence of Production Function on Capital or dependence of Consumption on Savings and so on. Different pairs of *self-conjugate* macroeconomic densities can have different forms of $Q_1$ and $Q_2$ factors. Let assume that dynamics of selected macroeconomic variable is determined by one *conjugate* variable only. Even for such simplification macroeconomic modeling remains extremely difficult. Origin of complexity concern diversity of possible forms of $Q_1$ and $Q_2$ factors in the right hand side of hydrodynamic-like equations (4,5). What does that mean for macroeconomic hydrodynamic-like models?

Let propose that right hand side factors $Q_1$ and $Q_2$ take form of simple linear operators on *conjugate* variables. Continuity Equations (4) have linear *scalar* right hand side factors $Q_1$ that depend on density $U$ or velocity $v$ of conjugate variable as:

$$1.\ Q_1 \sim U\ ;\ \ 2.\ Q_1 \sim \frac{\partial}{\partial t} U\ ;\ \ 3.\ Q_1 \sim \nabla \cdot v\ ;\ \ 4.\ Q_1 \sim \Delta U \tag{10.1}$$

Equations of Motion (5) have linear *vector* right hand side factors $Q_2$ that depend on density $U$ or velocity $v$ of conjugate variable as:

$$1.\ Q_2 \sim v\ ;\ 2.\ Q_2 \sim \frac{\partial}{\partial t} v\ ;\ 3.\ Q_2 \sim \nabla U\ ;\ 4.\ Q_2 \sim rot\ v\ ;\ 5.\ Q_2 \sim \Delta v \tag{10.2}$$

Relations (10.1) describe four possible scalar linear operators on *conjugate* density and velocity. Relations (10.2) present five *vector* linear operators on *conjugate* density and velocity. These linear operators describe *local* action of *conjugate* variables due to Eq.(4,5). For example Continuity Equation on macroeconomic density $U_M$ can have $Q_1$ factor that is proportional to *conjugate* density $U$ or proportional to time derivation of density $U$ etc. Let assume that different macroeconomic densities can depend upon *conjugate* variables in a different manner and present simple possible operators. It is obvious, that any linear composition of these operators can be used as a model for mutual dependence between macroeconomic variables on e-space. Usage of each possible form of $Q_1$ and $Q_2$ factors requires economical consideration and validation. Moreover, two *self-conjugate* macroeconomic variables may depend upon each other in a different manner. For example, variables $U_2$ may define $Q_1$ factor for Continuity Equation on variable $U_1$ as $Q_1 \sim \partial U_2/\partial t$ and variable $U_1$ may define $Q_1$ factor for Continuity Equation on variable $U_2$ as $Q_1 \sim \Delta U_1$. That increases diversity of different models of two *conjugate* macroeconomic variables interaction up to 200 versions. To describe real interaction between macroeconomic variables one should take into consideration action of two, tree or more *conjugate* variables. To develop model equations in a closed form one need to consider a system of tree, four or more hydrodynamic-like equations like Eq.(4,5). Different macroeconomic variables may have different forms of $Q_1$ and $Q_2$ factors and different substitutions of (10.1) and (10.2) relations make diversity of macroeconomic models incredibly huge.

Let outline some vital distinctions between economic and physical systems on one hand and let mention extremely interesting and tough problems to be solved by methods of statistical physics to establish economic theory in a rigorous way.

1. Economic space is defined by methods that adopt current economic phenomena's. It is impossible to define risk assessment on "empty economic space" without economic agents. Thus economics, at least in our model, has no analogies like free space, point particle mechanics, "fundamental" equations, conservation laws, symmetries and etc. It seems that economic theory begins with description of random multi-agent system. That is completely different from physics foundations. It seems that lack of economic conservation laws leads to lack of economic equilibrium states. That arises a lot of problems: How to develop non-equilibrium stationary states model for system that consists of many interacting subsystems and each particular subsystem does not have it's own equilibrium state? How to develop theory starting with kinetic-like description? How to develop kinetic-like and hydrodynamic-like theory without models and equations that describe dynamics of separate e-particles?

2. Economic agents or e-particles as we call them are completely different from physical particles. E-particles have size that can be determined by probability distribution that estimate risk ratings or coordinates of particular economic agent on e-space $R^n$. But e-particles do not collide with each other. Any number of different e-particles can exist in the neighborhood of point $x$ on e-space. No collisions between e-particles mean lack of economic analogy o pressure and viscosity. As we proposed in Sec. 4 for *local model* economic variables of e-particles depend on *conjugate* economic variables of different e-particle. Demand of e-particle do not depend directly on Demand of other particles, but on any other *conjugate* variables as Income, Saving, Supply and etc. It seems reasonable that set of e-particles can establish some stationary non-equilibrium state determined by interaction of *conjugate* variables. That arises problems: How to describe models for thermodynamic-like stationary state of multi-agents system on e-space. Such stationary states should be different for different set of interacting *conjugate* economic variables. How these states can interact with each other?

3. Economic analogy of kinetic distribution function helps develop transition from economic kinetic-like description to economic hydrodynamic-like description and that arises questions: How to derive economic kinetic-like equations on distribution functions without underlining equations of e-particles "mechanics"? How possible kinetic-like equations on distribution functions determine right-hand side factors of hydrodynamic-like equations? Do equations on distribution functions depend on *conjugate* variables?

That is only negligible number of problems that should be solved to establish some rigorous scheme for economic theory on economic space.

## 8. Conclusions

Introduction of economic space opens doors for wide usage of mathematical and statistical physics methods for economic modeling. Economic space employers risk ratings methods that were developed for decades. Risks should be treated as drivers of economic evolution and absence of any risks delete reasons for economic growth. Reasonable economic space should be determined by set of major risks that define current economic evolution. Economic space has different representations for different economic conditions under action of different major risks. It is important to develop procedures that can compare influence of different risks and can chose major risks that define economic space. That may allow compare predictions of macroeconomic models with observed macroeconometric data and may help establish econometric measurements *alike to* measurements in physics. We assume that it is impossible establish determined macroeconomic description. Random nature of risks growth and decline and random nature of economic space representations insert internal stochasticity into economic evolution and forecasts. Long-term macroeconomic forecasts require development of macroeconomic dynamics on current economic space $R^n$ and assumptions on future $m$ risks configuration that will define economic space $R^m$ in projected time term.

Modeling on economic space uncovers extreme complexity of macroeconomics even for simplest models in the assumptions of *local* interaction between economic agents on economic space. This assumption allows derive hydrodynamic-like equations on macroeconomic variables in a closed form. Incredible diversity of relations between macroeconomic variables on economic space allows develop models of mutual dependence that reflect specific economic nature of particular problem. Economic space gives ground for wide usage of mathematical and statistical physics methods and models. Differences between nature of economics and physics are so vital that leave no chance for direct application of physical methods and models. As well physical schemes and concepts like kinetics and hydrodynamics might be useful for economic theory. Macroeconomic wave equations for simplest Investment-Interest Rate interaction model uncovers existence of wide range of wave processes in macroeconomics and might be useful for crises forecasting. Even simple models of macroeconomic waves allow describe irregular time fluctuations of variables of entire economics. We believe that theory of economic waves could be very important for economic modeling and forecasting.

Economic theory on economic space requires appropriate econometric foundations. To develop reasonable model on economic space one should solve many methodological and econometric problems. Definition of economic space requires cooperative efforts of Central Banks, Rating Agencies, Economic and Finance Research Communities, Regulators, Statistical Bureaus, and Business etc. Many problems should be solved to establish appropriate econometric models on economic space. It is obvious that macroeconomic models can be developed on continuous spaces as $R^n$ and on discreet lattice as well. Lattice macroeconomic models require less changes of risk rating methodologies and can be developed within current risk ratings definitions.

## Acknowledgements

This research did not receive any specific grant from TVEL or funding agencies in the public, commercial, or not-for-profit sectors and was performed on my own account.